\documentclass[twocolumn,10pt]{article}

\usepackage[numbers]{natbib}

\usepackage[font={small}]{caption} 

\usepackage{graphicx}
\graphicspath{{./figures/}}

\usepackage{anysize}

\usepackage{amsmath}
\newcommand{\ud}{\,\mathrm{d}}
\numberwithin{equation}{section}

\usepackage{authblk}

\title{Guiding microscale swimmers using teardrop-shaped posts}

\author[1]{Megan S.\ Davies Wykes\thanks{megan.davieswykes@cantab.net}}
\author[2]{Xiao Zhong}
\author[1]{Jiajun Tong}
\author[2]{Takuji Adachi}
\author[1,3]{Yanpeng Liu}
\author[1]{Leif Ristroph}
\author[2]{Michael D.\ Ward}
\author[1,4]{Michael J.\ Shelley}
\author[1,5,6]{Jun Zhang}

\affil[1]{\small Applied Mathematics Laboratory, Courant Institute, New York University.}
\affil[2]{Molecular Design Institute, Department of Chemistry, New York University.}
\affil[3]{Institute of Fluid Mechanics, Beijing University of Aeronautics and Astronautics.}
\affil[4]{Flatiron Institute, Simons Foundation.}
\affil[5]{Department of Physics, New York University.}
\affil[6]{NYU-ECNU Institutes of Mathematical Sciences and Physics Research, NYU-Shanghai.}

\begin{document}

\twocolumn[
\begin{@twocolumnfalse}

\maketitle
\vspace*{-0.8cm}

\begin{abstract}
	The swimming direction of biological or artificial microscale swimmers tends to be randomised over long time-scales by thermal fluctuations. Bacteria use various strategies to bias swimming behaviour and achieve directed motion against a flow, maintain alignment with gravity or travel up a chemical gradient. Herein, we explore a purely geometric means of biasing the motion of artificial nanorod swimmers. These artificial swimmers are bimetallic rods, powered by a chemical fuel, which swim on a substrate printed with teardrop-shaped posts. The artificial swimmers are hydrodynamically attracted to the posts, swimming alongside the post perimeter for long times before leaving. The rods experience a higher rate of departure from the higher curvature end of the teardrop shape, thereby introducing a bias into their motion. This bias increases with swimming speed and can be translated into a macroscopic directional motion over long times by using arrays of teardrop-shaped posts aligned along a single direction. This method provides a protocol for concentrating swimmers, sorting swimmers according to different speeds, and could enable artificial swimmers to transport cargo to desired locations.\end{abstract}

\end{@twocolumnfalse}
]

Swimming at the microscale is randomised by thermal noise. Over long time-scales, micro-swimmers lose memory of their original direction due to rotational diffusion. As a result, microscale swimmers exhibit enhanced isotropic spatial diffusion rather than directed migration \cite{Howse2007,Palacci2010}. If microscale swimmers are to be employed for useful tasks such as cargo carrying, \cite{Burdick2008,Wang2012a} new approaches for directing their motion are required. Macroscopic directed motion in natural systems is usually achieved through response to external fields, such as a flow field \cite{Bretherton1961,Marcos2012}, gravitational field \cite{Pedley1992}, or chemical gradient \cite{Berg1972,Mitchell2006,Friedrich2007}. In contrast, hydrodynamic interactions between artificial micro-swimmers and the surfaces of obstacles offer a new possibility for locally guiding microscale swimmers without the use of an externally imposed field.

\begin{figure}[h!]
\centering
	\includegraphics[width=0.45\textwidth]{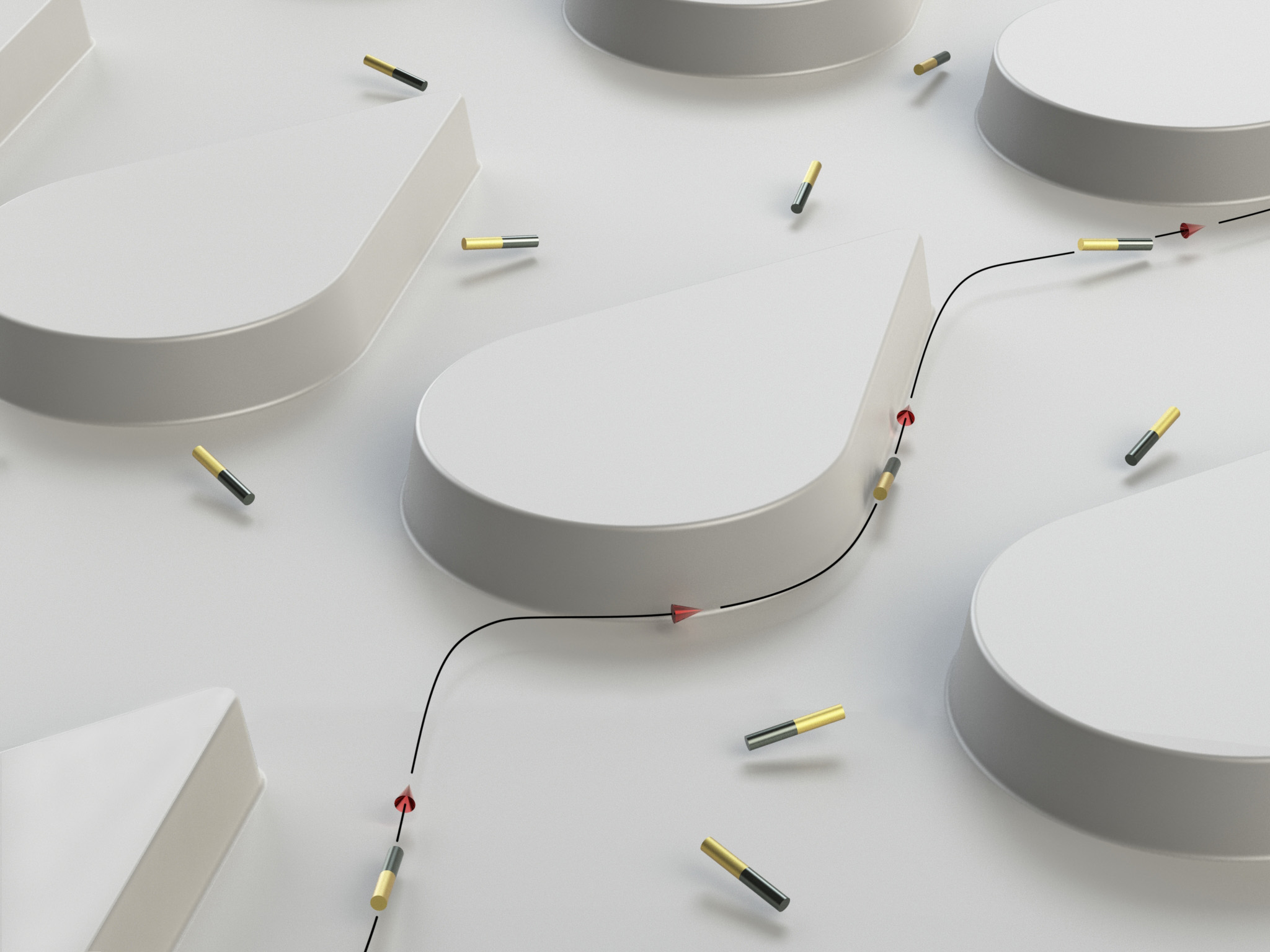}
	\caption{Sketch of artificial Au-Pt nanorod swimmers interacting with teardrop-shaped posts. Rods swim with the Pt-end leading when placed in a solution of hydrogen peroxide. After encountering a post, swimming rods tend to travel in proximity to the perimeter, then preferentially depart from the tip where the curvature is highest.}
	\label{fig:teardrop_diagram} 
\end{figure}

Higher concentrations of microswimmers near surfaces have been attributed to a combination of hydrodynamic  and  steric  interactions \cite{Berke2008,Li2009,Contino2015,Sipos2015}. When \emph{E.\ coli} bacteria or \emph{Chlamydomonas} algae encounter a surface, they are deflected at a small scattering angle that is independent of the angle of incidence \cite{Galajda2007,Kantsler2013}. Using a wall of funnels or ratchets, this effect has been used to concentrate suspensions of bacteria or algae \cite{Galajda2007,Kantsler2013,Lambert2010}.

Although the swimming patterns of artificial microswimmers are arguably simpler than their natural counterparts, they can exhibit complicated behaviour near surfaces of obstacles. For example, nanorod swimmers consisting of conjoined gold and platinum ends can be captured by and orbit spheres encountered in their paths before escaping due to thermal fluctuations \cite{Takagi2014}. Artificial Janus particle swimmers composed of polystyrene colloids, sputtered on one side with platinum, can `hop' between spheres in a crystalline array of polystyrene colloids, with the hopping rate dependent on the fuel concentration, but independent of the swimming speed \cite{Brown2016}. In another report, Janus swimmers of a similar type were found to orbit circular posts. Under high fuel concentration there is a suggestion that the residence time of swimmers (i.e.\ the amount of time a swimmer followed the surface of a post) increased with increasing diameter of circular posts \cite{Simmchen2016}. Collectively, these observations suggest a new approach for directing the motion of artificial microswimmers based on asymmetric posts with variable curvature along their circumference, as it is to be expected that swimmers would leave faster from regions with higher curvature than from regions with lower curvature. A teardrop shape was chosen as it is a simple shape that concentrates curvature at one end, allowing for more effective retention at the blunt end and detachment at the sharp tip.

Herein, we explore this approach using gold-platinum segmented rods that swim in aqueous media containing hydrogen peroxide fuel. These artificial swimmers are captured by teardrop-shaped posts, swim in close proximity to the post perimeter, then preferentially depart from the high curvature end (Fig.\ \ref{fig:teardrop_diagram}). This produces a bias in the direction of motion that is expressed macroscopically when aligned arrays of these posts are used. Swimming rods are then more likely to cross between rows of teardrops in an array of teardrop-shaped posts in the direction in which the teardrops are pointing. Herein, it is shown that the effect depends on swimming speed of the rods and the spacing between posts in the array. 

\section{Experiments}

The artificial swimmers used in this study are 300 nm diameter nanorods consisting of two solid segments, one gold (Au) and one platinum (Pt), fabricated by electrochemical deposition in anodized aluminum oxide (AAO) membranes \cite{Banholzer2009,DaviesWykes2016}. The rods are $2.2 \pm 0.4 \, \mu$m in length. These bimetallic rods swim by self-electrophoresis when placed in a solution containing hydrogen peroxide (H$_2$O$_2$) fuel \cite{Paxton2004,Moran2010,Paxton2006,Takagi2013,Takagi2014,Wang2006,Wang2009,Ebbens2010,Wang2013a}. Electrochemical decomposition of H$_2$O$_2$ results in a gradient in proton concentration, which corresponds to an electric field pointing from Pt to Au. The rods themselves have an overall negative charge. Therefore, the positively charged electrical double layer surrounding the rod experiences a force due to the self-generated electrical field. A fluid flow develops on the rod surface, from Pt to Au, causing the rod to swim with its Pt end leading \cite{Paxton2006,Moran2010,Moran2011,Wang2013a}. This mechanism for creating microscale flows has been used in several previous studies on artificial active particles \cite{Paxton2004,Paxton2006,Wang2006,Wang2009,Ebbens2010,Takagi2013,Takagi2014,Wang2013a,Wang2015}. A video of these nanorods swimming is provided as supplementary video S1, revealing strong random fluctuations in swimming direction. 

The rods are dense compared to the solution, so they swim on the surface of a microscope coverslip or glass wafer, resulting in a system which is quasi-2D. The velocity of each rod is measured from its particle track using the mean squared displacement\cite{Dunderdale2012}, using tracks observed over a minimum of 100 frames, at a frame rate of 10 fps. In typical experiments using glass wafers, a wide distribution of rod speeds is observed. For example, at a fuel concentration of 10\%, rods are observed to swim at speeds between 0 and 5 microns/s. Such dispersion in swimming speed is possibly due to differences in the individual rods created during their fabrication. Measurements found a weak negative correlation between rod length and swimming speed (correlation coefficient, $\rho =  -0.41$), suggesting longer rods swim more slowly. 

\begin{figure}[t]
	\centering
	\includegraphics[height=3.4cm]{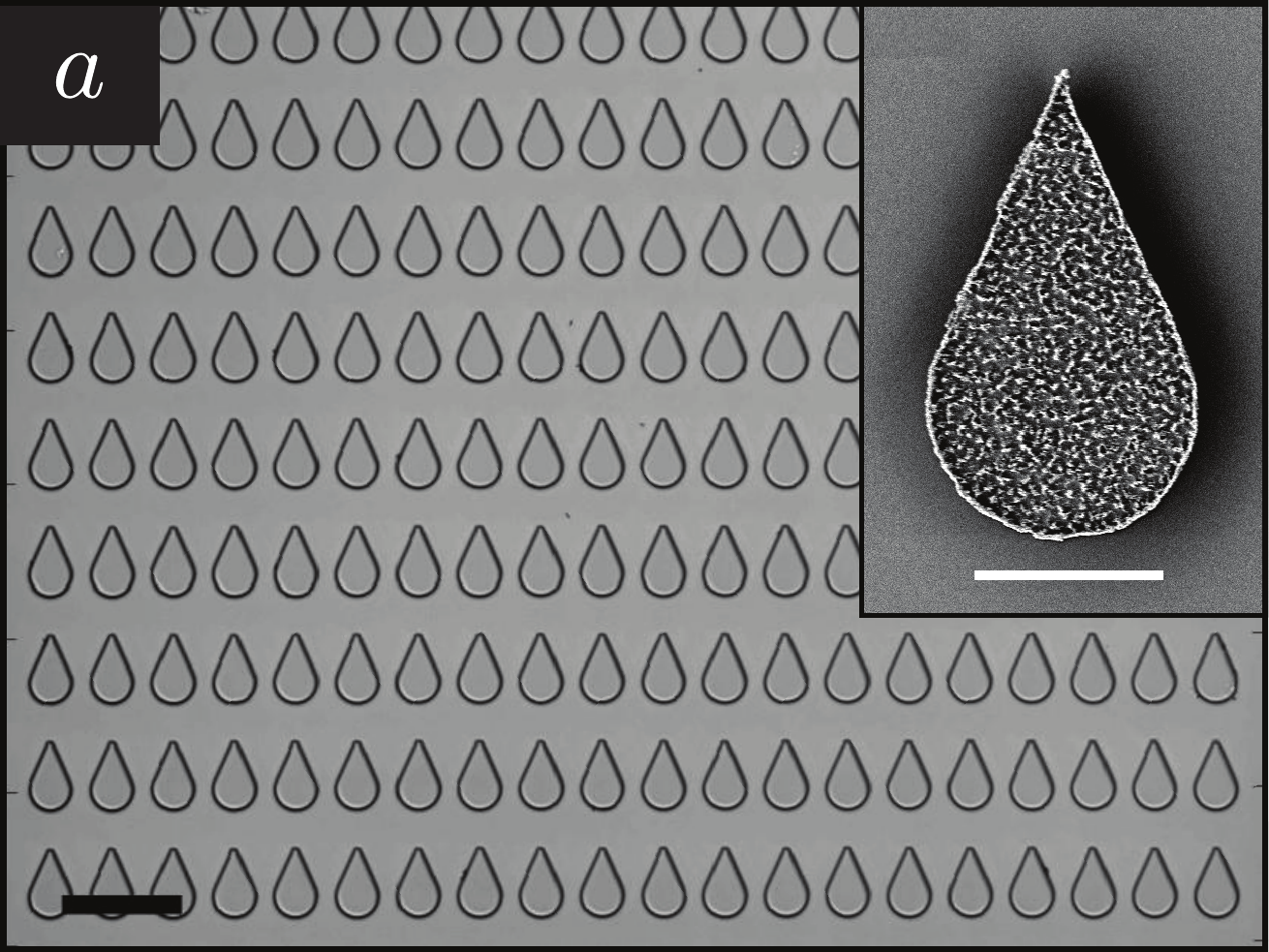}~~
	\includegraphics[height=3.4cm]{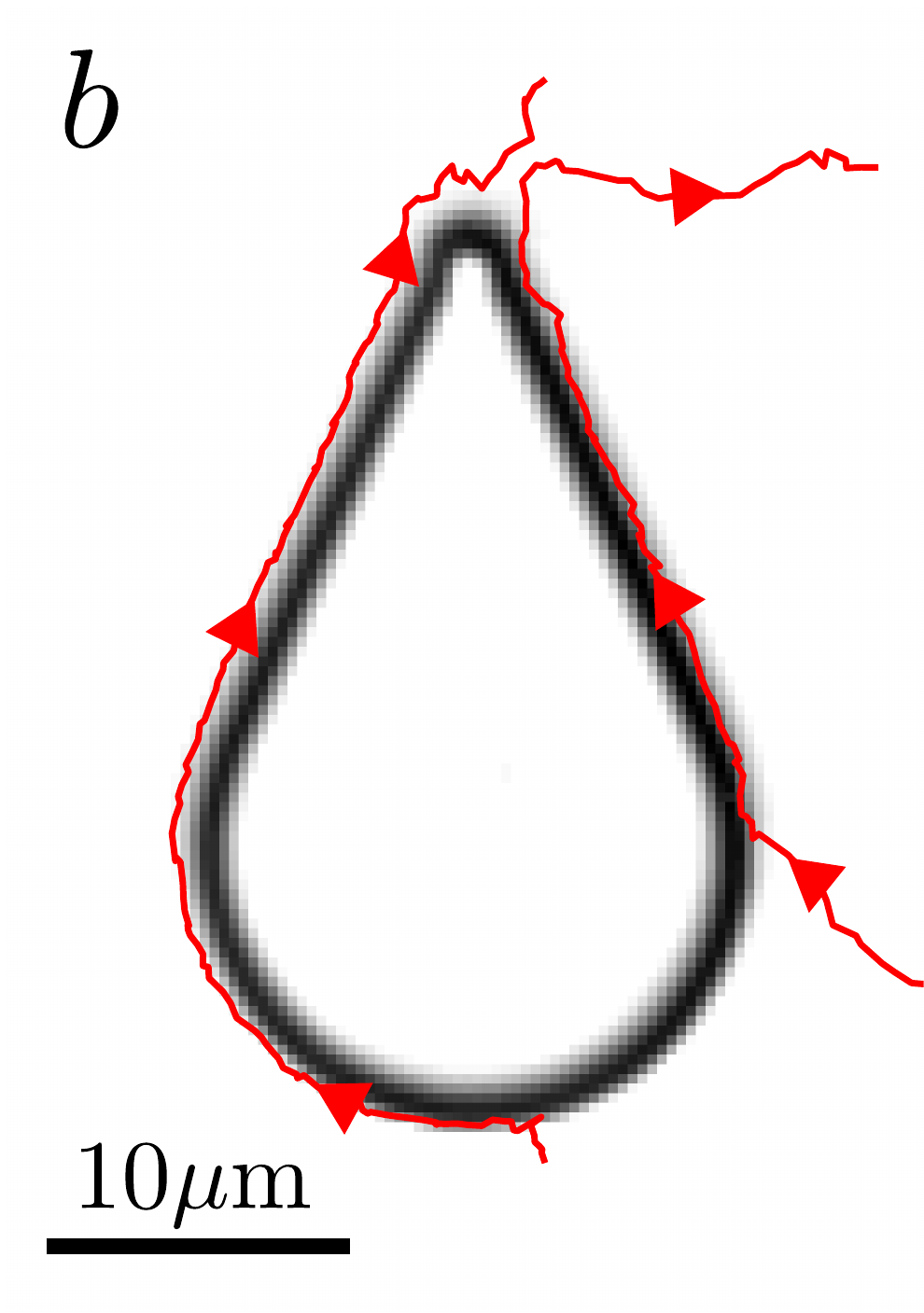}
	\vspace*{0.3cm}
	\caption{Experiments are performed using arrays of teardrop-shaped posts and artificial microscale swimmers: (a) Photograph of teardrop post array (scale bar $50 \, \mu\text{m}$). Posts are 30 $\mu$m in length, $4\mu$m tall, with gaps of $15 \, \mu \text{m}$ between rows and $7.5 \, \mu \text{m}$ between columns. Inset: Scanning electron microscope (SEM) image of a single teardrop post (scale bar $10 \, \mu\text{m}$). (b) Example tracks of artificial swimmers (not shown) interacting with a teardrop-shaped post.} \label{fig:teardrop_SEM}
\end{figure}

Teardrop-shaped posts (Fig.\ \ref{fig:teardrop_SEM}a) were patterned onto borosilicate glass wafers (University Wafer) using SU-8 2005 negative photoresist (Microchem Corporation). The photoresist was spun cast at $500$ rpm for $10$ s, then at $3000$ rpm for $30$ s, onto an air plasma-treated wafer and the wafer was subsequently baked at $95^\circ$C for $2$ min. After cooling, the wafer assembly was exposed to $360$ nm UV light by a mask aligner (Karl Suss MJB3) at an intensity $3.22$ mW/cm$^2$ for $32$ s through a photomask (DigiDat), on which desired patterns were printed. The exposed wafer assembly was then baked again at $95^\circ$C for $3$ min and cooled to room temperature before submerging in SU-8 developer for $1$ min to remove unexposed photoresist. Then the wafer, now with arrays of tear-dropped posts, was rinsed with fresh SU-8 developer and isopropyl alcohol and dried with air. Experiments are then performed directly on the printed wafer to preserve the quality of the posts (Fig.\ \ref{fig:teardrop_SEM}a, inset).

The posts are 30 $\mu$m long and approximately 4 $\mu$m tall, as measured with a scanning electron microscope. The perimeter of a post is 77 $\pm$ 2 $\mu$m. Two arrays were used, with the same teardrop shape for each array, but different spacings between posts. Most experiments were performed using a loosely packed array, which had gaps of 15 $\mu$m between rows of posts and 7.5 $\mu$m between columns (Fig.\ \ref{fig:teardrop_SEM}a). A densely packed array was also used, with gaps of 7.5 $\mu$m between rows and 3.75 $\mu$m between columns.

Images from experiments were collected using an inverted optical microscope and a $20\times$ air objective. An O-ring with diameter 5 mm was placed on top of the glass wafer printed with teardrop-shaped posts. The region confined by the O-ring was filled with 20 $\mu$l of a dilute suspension of bimetallic rods in aqueous hydrogen peroxide solution. A glass cover slip was then placed over the top of the O-ring to reduce evaporation. The O-ring contains the solution of fuel and rods and was made of rubber, which was chosen to avoid damaging the surface of the printed wafer. Videos of experiments were recorded at 10 frames per second for 10 - 20 minutes. The concentration of rods was sufficiently low to prevent significant fuel depletion and a corresponding reduction in average velocity of rods.

\section{Results}

\begin{figure}[t]
	\includegraphics[width=0.45\textwidth]{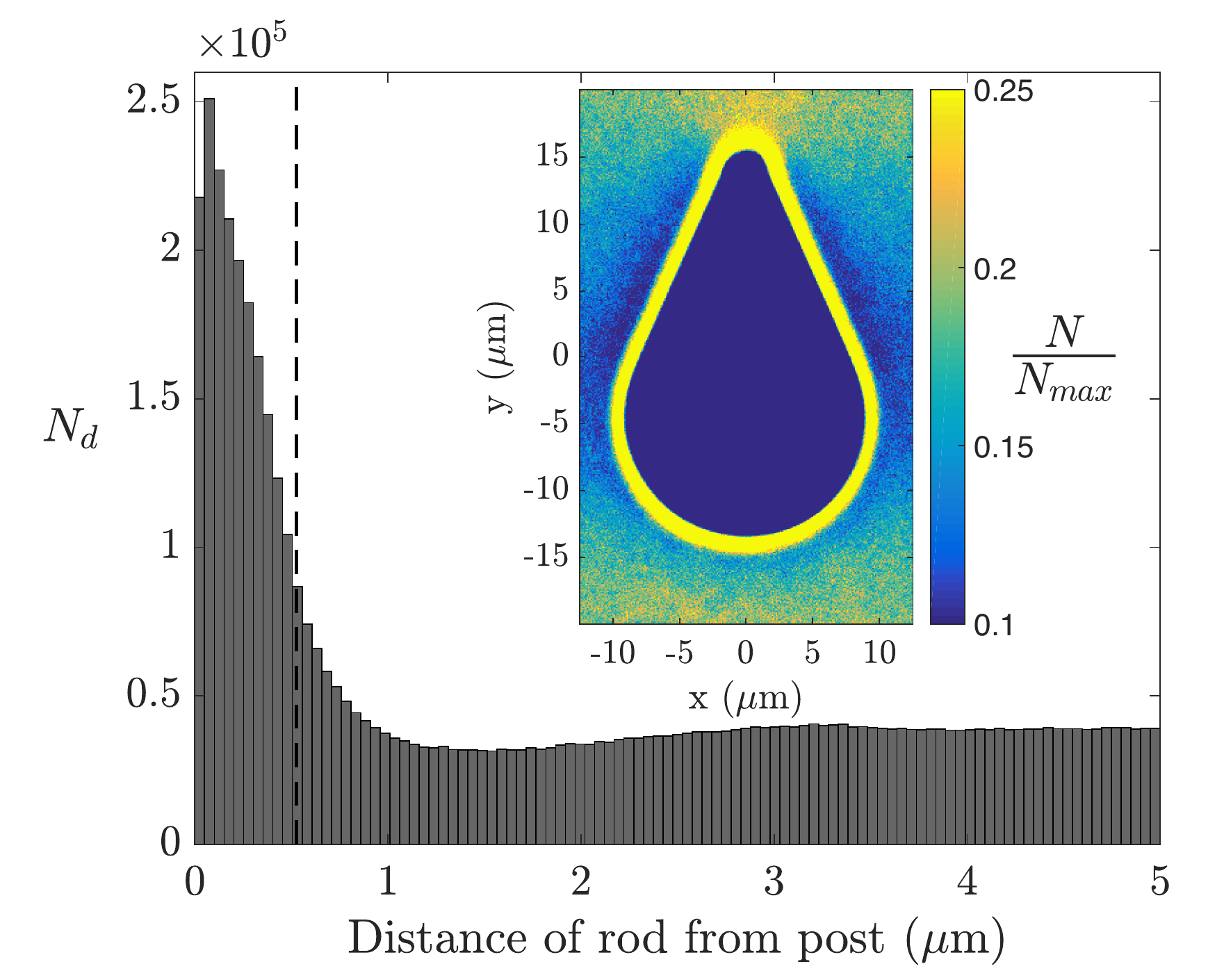}
	\caption{Rod locations relative to posts: Histogram of distance of rod centroids to the edge of the nearest post. Threshold for when a rod is considered to be on/off teardrop is $1/e$ of maximum frequency and is shown as dashed line. Inset: 2D histogram of relative rod concentration around a teardrop-shaped post, with vertical and horizontal bin widths of $0.1 \, \mu$m, calculated using $9 \times 10^6$ data points from 11 experiments using the array shown in Fig.\ \ref{fig:teardrop_SEM}a. Colour indicates number of rods seen at that position, divided by the maximum number of rods observed in any position.} \label{fig:distance_from_post}
\end{figure}

The nanorod swimmers are attracted to the surface of the posts, swimming alongside the post perimeter, and then departing at some point on the surface (see supplementary movies S2, S3). Two example tracks of swimmers interacting with a teardrop-shaped post are shown in Fig.\ \ref{fig:teardrop_SEM}b. A few rods were observed to perform several complete orbits around a post. A histogram of the distance of the rod centroid from the surface of the closest post is shown in Fig.\ \ref{fig:distance_from_post}. Swimmers are concentrated in a thin region close to the post perimeter; a slight depletion is observed around $1-2 \, \mu \text{m}$ from the post. The concentration is seen to be uniform outside this region, consistent with previous results \cite{Takagi2014}.

The spatial distribution of rods around a post is examined further by determining the relative position of the centroid of each rod with respect to the closest teardrop in an array. The 2D histogram of this ensemble is plotted in Fig.\ \ref{fig:distance_from_post} (inset), where the colour corresponds to the number of rods at that location divided by the maximum number of rods in any location. This analysis reveals that on average there is an increased concentration of rods around the tip and slightly below the base of a post, a decreased concentration of rods near the flat sides, and a high concentration of rods on the perimeter (Fig.\ \ref{fig:distance_from_post}).

The behaviour of the nanorod swimmers is analysed by examining joining and leaving events along a post perimeter. A swimmer is considered to have left a post if it is within a threshold distance from the post perimeter for at least two consecutive frames ($0.2$s) and then remains beyond the threshold distance for at least two consecutive frames. A joining event is defined in a similar way, in the reverse order. The threshold distance is chosen as that where the frequency falls below $1/e$ of the maximum frequency in the histogram of rod distances from the nearest post (Fig.\ \ref{fig:distance_from_post}, dashed line). 

\begin{figure}[t]
	\centering
	\includegraphics[width=0.45\textwidth]{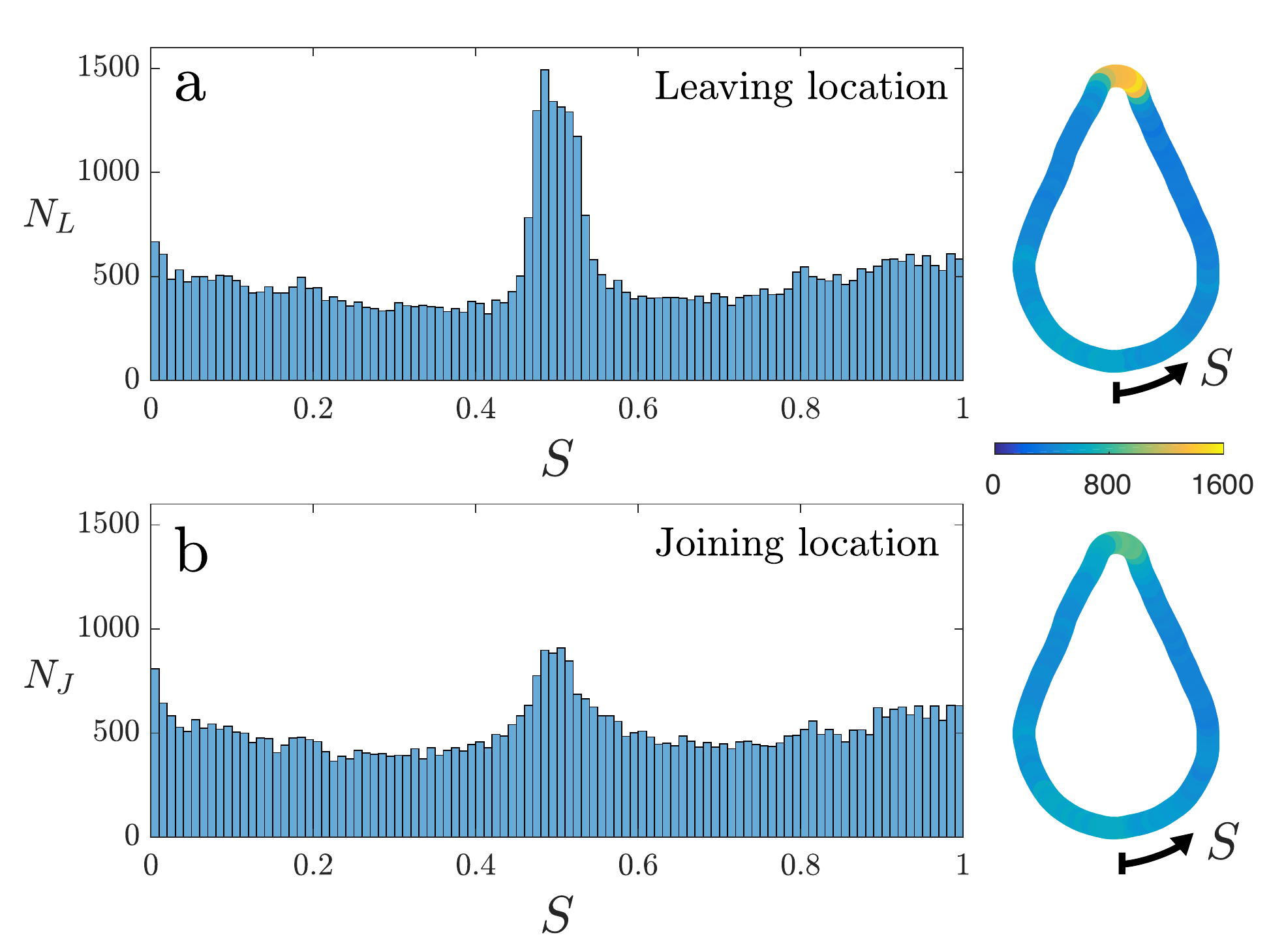}
	\caption{Leaving and joining locations around a teardrop: Histogram of (a) leaving events and (b) joining events as a function of arc-length $S$. Data from 13 experiments using arrays of teardrop-shaped posts and rods with a swimming speed $> 1 \mu$m/s. Outlines on the right show the number leaving (or joining) at arc-length $S$, plotted around a teardrop.}
	\label{fig:hist_leave_join}
\end{figure}

The dependence of the frequency $N$ of leaving and joining events on arc-length $S$ is illustrated in Fig.\ \ref{fig:hist_leave_join}, where $S$ has an origin at the base of the post circumference and continues in the counter-clockwise direction (Fig. 4a and b). These histograms combine the results of 11 experiments, with the locations of more than $10^4$ leaving and joining events from posts in an array. Fig.\ \ref{fig:hist_leave_join}a reveals that the swimmers are more likely to leave the post near the sharp tip (around $S = 0.5$). There is also some variation in the histogram of locations of joining events (Fig.\ \ref{fig:hist_leave_join}b), with more joining events close to the tip ($S = 0.5$) and base ($S$ near $1$ or $0$) of a post.

Experiments are performed using arrays of teardrops, which could explain the variation in joining frequency along the post perimeter. Since more rods leave at the tip of the teardrop, on average there is an increased rod concentration in the region close to the tip. This is confirmed by the average spatial distribution of rods around a post (Fig.\ \ref{fig:distance_from_post}, inset). Due to the randomisation of the swimming direction by thermal noise, rods can return and re-join the tip of the teardrop. As the posts are in an array, the tip of one teardrop post is close to the base of another. Hence, the increased average concentration of rods around the tip of one post corresponds to an increased average concentration near the base of a post in the adjacent row. This would result in more rods joining at the base of a post.

\begin{figure}[t]
	\includegraphics[width=0.45\textwidth]{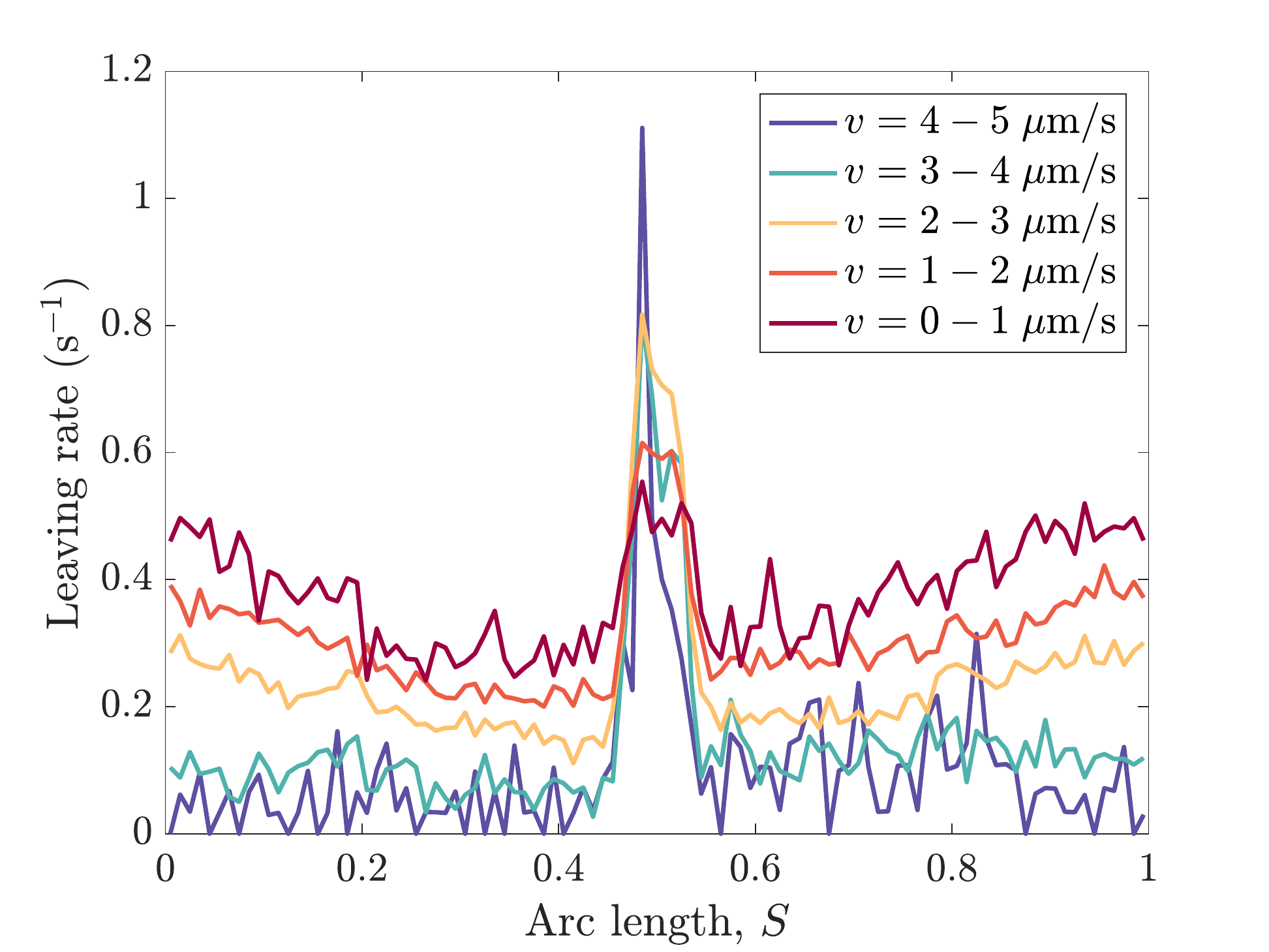}
	\caption{Interaction of a swimming rod with a single teardrop: Leaving rate from a teardrop with arc-length $S$. Calculated from the number of rods that leave at an arc-length (Fig.\ \ref{fig:hist_leave_join}a), divided by the number of times a rod was observed at that arc-length.} \label{fig:single_teardrop_leaving_rate}
\end{figure}

To identify whether there is a variation in the rate at which rods leave the post perimeter, the leaving rate for each section of arc-length is determined by dividing the frequency of leaving events in that section by the average number of rods present in that section. This is plotted as functions of $S$ for rods at different speeds in Fig.\ \ref{fig:single_teardrop_leaving_rate}. When rods swim along the perimeter, their swimming speeds do not change significantly compared with their free-space speeds\cite{Takagi2014}. Consequently, the swimming speed on the perimeter is not distinguished from the swimming speed away from a post.

For rods swimming at speeds less than $1 \, \mu \text{m}/s$, there is already a small contrast between the leaving rates at the tip and the flat surfaces of the teardrop-shaped post (Fig.\ \ref{fig:single_teardrop_leaving_rate}). As the swimming speed increases, the tendency of rods to leave at a higher rate from the tip of the teardrop becomes more significant; the contrast between the leaving rates  at the tip and the sides of a teardrop increases monotonically. This is consistent with previous results for Janus particles orbiting circular posts, wherein the residence times of swimmers on the perimeters of circular posts were longer for posts with a larger radius \cite{Simmchen2016}. There were fewer leaving events for the highest velocity window ($4 - 5 \, \mu \text{m}/s$); therefore, the curve displays more variability than those at lower velocities.

\begin{figure}[t]
	\includegraphics[width=0.45\textwidth]{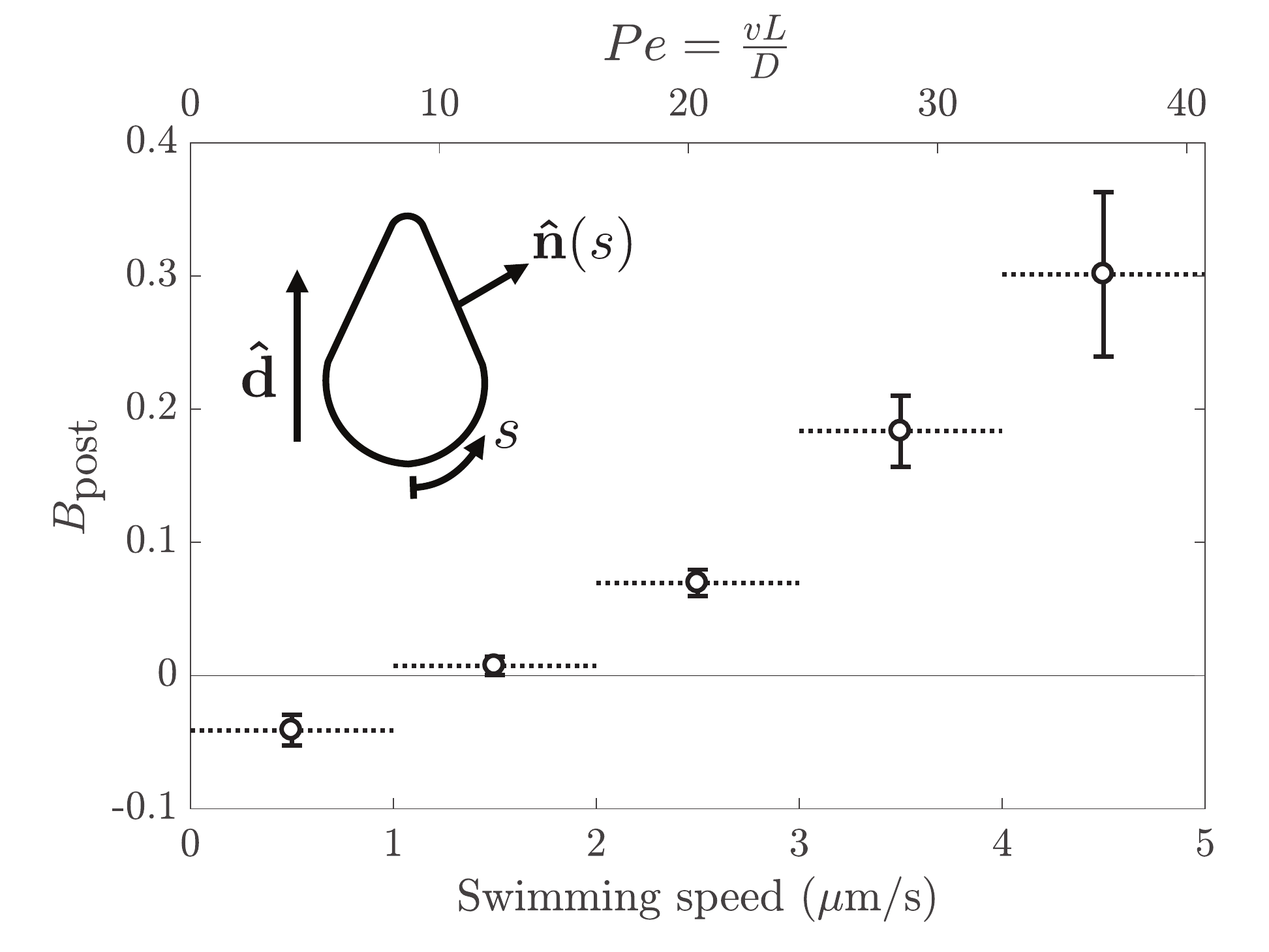}
	\caption{Post bias (defined by Eq.\ \ref{eq:local_bias}) as a function of swimming speed. A positive post bias indicates rods are preferentially leaving from a region of the post with a surface normal in the direction $\textbf{\^{d}}$. Dotted horizontal lines correspond to range of rod velocities used to calculate each datapoint, vertical solid lines show one standard deviation. The Peclet number $Pe = vL/D$, where $v$ is the deterministic swimming speed of a rod, $L$ is the rod length and $D$ is the translational diffusion coefficient.} \label{fig:single_teardrop_bias}
\end{figure}

The overall effect of an individual teardrop-shaped post on the motion of rods is characterised by a metric denoted here as the \emph{post bias}. The post bias depends on the distribution of leaving events and the orientation of the surface from which rods leave, compared with the overall orientation of the teardrop. The post bias is defined as
\begin{equation}
	B_{\text{post}} = \frac{\int N_{L}(S) \cdot \textbf{\^{d}} \cdot \textbf{\^{n}}(S) \ud S}{\int N_{L}(S) \ud S}, \label{eq:local_bias}
\end{equation}
where $N_{L}(S)$ is the number of leaving events as a function of $S$, as shown in Fig.\ \ref{fig:hist_leave_join}a, $\textbf{\^{n}}(S)$ is a unit vector normal to the post perimeter at $S$, and $\textbf{\^{d}}$ is the unit orientation vector pointing in the direction of the teardrop (see Fig.\ \ref{fig:single_teardrop_bias}, inset). The surface normal is assumed to be a good approximation for the average direction in which a rod leaves the post perimeter. The post bias $-1 \le B_{\text{post}} \le 1$, with $B_{\text{post}} = 1$ corresponding to all rods leaving the post from a surface pointing in the direction of $\textbf{\^{d}}$ and $B_{\text{post}} = -1$ corresponding to all rods leaving the post from a surface pointing in the $-\textbf{\^{d}}$ direction.

The variation in $B_{\text{post}}$ with swimming speed is shown in Fig.\ \ref{fig:single_teardrop_bias}. For rods that are swimming slowly, there is a small overall bias in the orientation of the surface from which rods leave. As the velocity increases, swimmers are more likely to leave from a portion of the surface that points in the $\textbf{\^{d}}$ direction (such as the tip of the teardrop-shaped post). A possible interpretation is as follows. As the speed increases, the rate at which rods leave the perimeter of the post decreases everywhere except at the tip where it increases (Fig. \ref{fig:single_teardrop_leaving_rate}). This means that faster rods joining the post are likely to travel further along the perimeter of the post and reach the tip, where they are then more likely to leave. As faster rods have a greater chance of reaching the tip, the post bias increases.

\begin{figure}[t]
	\centering
	\includegraphics[width=0.45\textwidth]{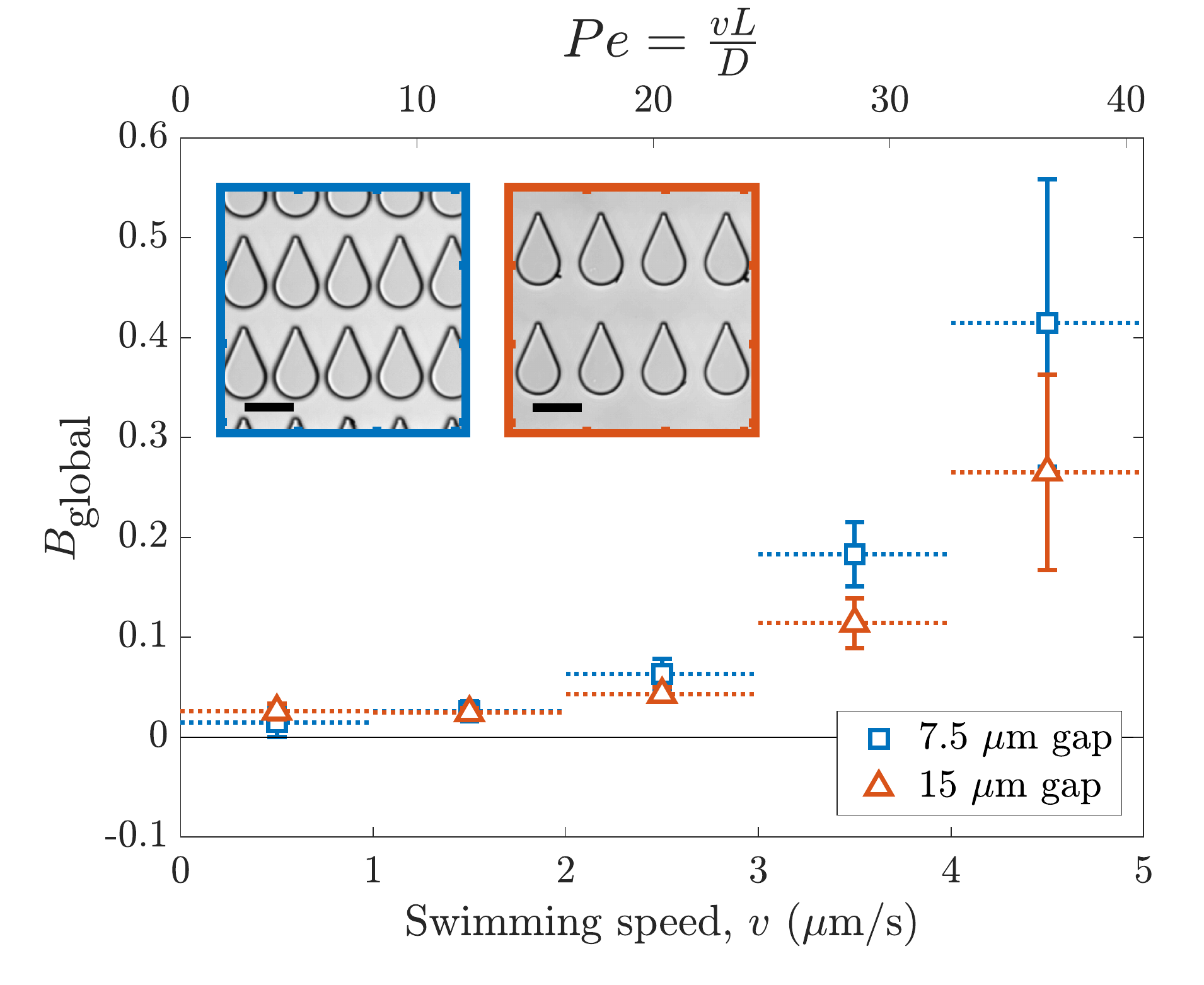}
	\caption{Global bias (defined by Eq.\ \ref{eq:global_bias}) as a function of swimming speed. A positive global bias indicates that as rods travel through an array they are more likely to be travelling in the direction $\textbf{\^{d}}$ (Fig.\ \ref{fig:single_teardrop_bias}). Dotted horizontal line corresponds to range of rod velocities used to calculate the datapoint, vertical solid lines show one standard deviation. Insets show images of each array type with a scale bar of $20\mu\text{m}$. The Peclet number $Pe = vL/D$, where $v$ is the deterministic swimming speed of a rod, $L$ is the rod length and $D$ is the translational diffusion coefficient.} \label{fig:array_bias}
\end{figure}

The Peclet number describes the ratio of advective motion to diffusive motion for a particle that is both swimming and diffusing. A Peclet number can be defined as $Pe = vL/D$, where $v$ is the deterministic swimming speed of a rod, $L$ is the rod length and $D$ is the translational diffusion coefficient. The $Pe$ is plotted on the top axis of Figs.\ \ref{fig:single_teardrop_bias} and \ref{fig:array_bias}.

The data acquired for an individual post demonstrate the introduction of a bias into the orientation of the surface from which rods leave a teardrop-shaped post. The next step is measuring the collective effect of multiple posts in an array on swimming direction. To do this the swimming direction of rods crossing virtual horizontal lines half-way between rows of posts is examined. The total number of rods crossing any of these lines in the $\textbf{\^{d}}$ direction is $N_{+}$ ; similarly, the total number of rods crossing any one of the lines in the $-\textbf{\^{d}}$ direction is $N_{-}$. The \emph{global bias} of the array is defined as
\begin{equation}
	B_{\text{global}} = \frac{N_{+} \, - \, N_{-}}{N_{+} \, + \, N_{-}}. \label{eq:global_bias}
\end{equation}
This bias $-1 \le B_{\text{global}} \le 1$, with $B_{\text{global}} = 1$ corresponding to all rods crossing between rows in the $\textbf{\^{d}}$ direction and $B_{\text{global}} = -1$ corresponding to all rods crossing in the $-\textbf{\^{d}}$ direction. The global bias was chosen as a statistically reliable ensemble measure of the net flux of swimmers through an array.

In Fig.\ \ref{fig:array_bias}, the variation of the global bias with velocity is plotted for a loosely packed array (triangles, orange) and a closely packed array (squares, blue). Both arrays produced a positive bias for rods swimming at speeds more than $2-3 \mu \text{m}/s$, consistent with an overall macroscopic preference for travel in the direction $\textbf{\^{d}}$. The global bias increases with rod speed. This may be attributable to an increase in the post bias with speed, as faster rods are more likely to have left a teardrop-shaped post from the tip.

Another factor that could account for the increase in global bias with swimming speed is an increase in the distance travelled by a micro-swimmer before the swimming direction is randomised by rotational diffusion. The distance a rod moves before it has forgotten its initial direction scales with $v/D_r$, where $v$ is the rod speed and $D_r$ is the rotational diffusion coefficient. If $D_r$ does not vary with $v$, faster rods would be expected to travel further before their direction is randomised. This would be consistent with the increase in the global bias with increasing swimming speed if $v/D_r$ is of the same order as the distance between posts, as rods are more likely to encounter the next row of posts if they leave from the top of a post.

The translational diffusion coefficient $D$ for a passive rod at infinite dilution and far from any boundaries is
\begin{equation}
	D = \frac{k_B T}{3 \pi \mu L} \log\left(\frac{L}{d}\right),
\end{equation}
where $L$ is the rod length, $d$ is its diameter, $k_B$ is the Boltzmann constant, $\mu$ is the solvent viscosity, and $T$ is temperature and end corrections have been neglected \cite{Eimer1991}. The rotational diffusion coefficient is directly related to the translational diffusion coefficient by \mbox{$D_r = 9D/L^2$}. For \mbox{$L = 2.2 \, \mu$m,} \mbox{$d = 0.3 \, \mu$m,} \mbox{$\mu = 1.004 \times 10^{-3}$ Pa s,} \mbox{$T = 273$ K,} \mbox{$k_B = 1.38 \times 10^{-23} $ m$^2$kg/s$^2$K,} the theoretical value is \mbox{$D = 0.39 \, \mu$m$^2/$s.} The rods in this study swim near the surface of the glass wafer, and a translational diffusion coefficient smaller than the free space value would be expected.\cite{Feitosa1991}

The value of $D$ measured for rods on a coverslip in the absence of hydrogen peroxide is $D = 0.27 \, \mu$m$^2/$s. The measured value of $D$ is unchanged upon introduction of hydrogen peroxide and the rods swimming. The value of $D$ was measured for swimming rods using the mean squared displacement measured for times smaller than $1/D_r$, the rotational diffusion time scale\cite{Dunderdale2012}. A translational diffusion coefficient of this value corresponds to a rotational diffusion coefficient $D_r = 0.50$ s$^{-1}$ for rods of length \mbox{$L = 2.2 \mu$m,}. The time required for a swimmer to lose its memory about the starting direction is of the order $\tau_r = 1/D_r = 2$ s. The distance $v/D_r$ for rods with velocities between $0$ and $5 \, \mu$m/s, will range from $0$ to $10 \, \mu$m and therefore at higher speeds is of the same order as the distances between teardrops for both the closely packed and loosely packed arrays. 

\section{Discussion}

The results shown above demonstrate that the swimming direction of artificial bimetallic swimmers is biased by their interactions with teardrop-shaped posts. We have previously argued for their attraction and adherence to surfaces based on the nature of fluid lubrication forces arising from their phoretic self-propulsion \cite{Takagi2014}, while other arguments for hydrodynamic attraction have used far-field force dipole models for swimmers \cite{Spagnolie2015}. In any case, the higher departure rate of swimmers from teardrop tips is likely due to an abrupt decrease in their attractive interactions with the teardrop surface as they reach these high curvature regions. Along the low curvature regions of the teardrop base and sides, a rod has good contact with the wall and fluid lubrication forces from self-propulsion are expected to lead to an attraction that maintains close contact \cite{Takagi2014}. At the high curvature region of the tip, a rod abruptly loses much of its near contact with the wall and its attraction to it is expected to decrease. The teardrop sides and base thus function as a large ``collector" that adheres swimmers in the near vicinity, the tip serves as an ``emitter" that releases rods, and the displacement between collector and emitter leads to the biased motion.

This reasoning can be supported by more quantitative estimates of the effective distances between the rod and post. When thermal energy $k_BT/2$ acts upon a nanorod of buoyant weight $mg$, its characteristic sedimentation height is $d_{sdmt} = k_BT/2mg$, which is about 40 nm in our experiment \cite{Takagi2013,Driscoll2016}. This is the average distance between a flat substrate and a nanorod that sits above it. If the nanorod is set in motion by the chemical fuel, an attractive lubrication force becomes operative and draws the nanorod closer to the substrate \cite{Takagi2014}. The actual gap separation to the substrate is thus reduced and so $d_{actual} < d_{sdmt}$. 

These attractive interactions are substantial as is evidenced by the observation that swimming nanorods adhere to the substrate even when the rod-substrate system is flipped upside down \cite{Liu2017}. When a rod interacts with a sidewall of a teardrop post, in the direction orthogonal to the gravitational pull, the gap between the two is set by balancing the lubrication force and the thermal energy. Thus, the resultant gap distance should be less than the sedimentation height $d_{sdmt}$.

The typical effective radius of the teardrop tip is on the order of $1 \, \mu$m, as estimated from scanning electron microscope (SEM) images (Fig. \ref{fig:teardrop_SEM}a, inset). When a 2-micron long rod passes around the tip, a simple estimate gives the average gap distance between the rod and the tip wall as $\sim 500$ nm. This is at least a 10-fold increase over the gap distance $d_{actual}$ when the rod is moving along a flat sidewall to the tip region. This increased gap distance should severely weaken the attractive interactions between the rod and tip, leading to the high rate of departure of swimming nanorod from teardrop tips.     

\begin{figure}[t]
	\centering
	\includegraphics[width=0.49\textwidth]{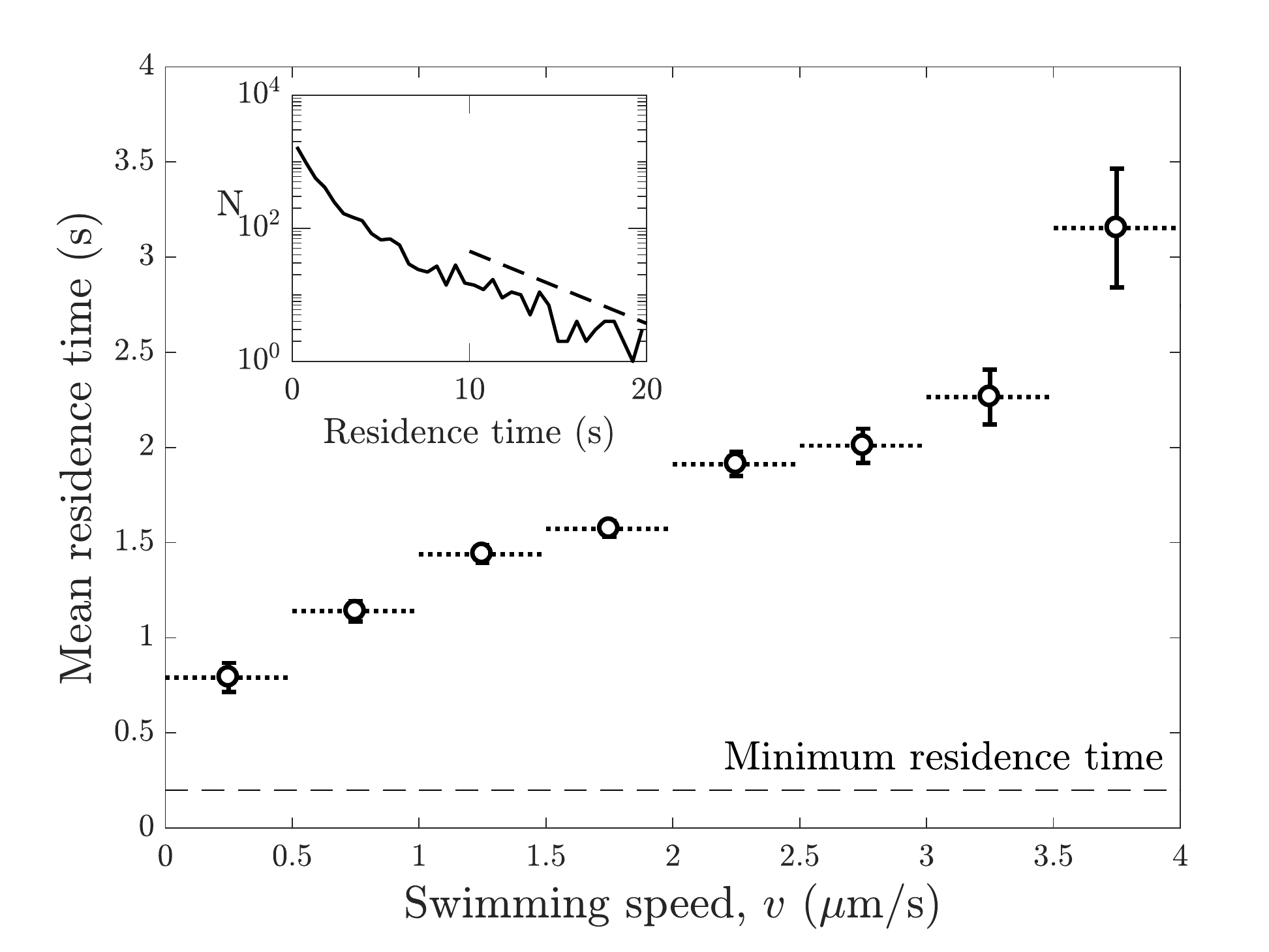}
	\caption{Mean time spent on teardrop-shaped posts as a function of mean swimming speed. Mean swimming speed calculated using entire swimmer path, both on and off post. The minimum length of time a rod must be close to a teardrop for it to be considered to have joined the drop is 0.2 s, this minimum residence time is indicated with a dashed line. All experiments had a H$_2$O$_2$ concentration of 10\%. Dotted horizontal line corresponds to range of rod velocities used to calculate the datapoint, vertical solid lines show the standard error of the mean. Inset shows histogram of residence times for rods swimming between $2 - 4 \mu$m/s. Dashed line has slope of $-0.25$.}
	\label{fig:time_on_teardrop}
\end{figure}

More information about the interaction between rods and teardrop-shaped posts can be gleaned from the mean residence time of a swimmer on a post. Measurements of the mean residence time for swimmers with different swimming speeds (Fig. \ref{fig:time_on_teardrop}) demonstrate that the mean residence time increases with increasing swimming speed, which is consistent with the lubrication model \cite{Takagi2014} and a pusher force-dipole model \cite{Spagnolie2015}. There is some length polydispersity in our swimmer population, and longer rods might be expected to have longer residence times as the rotational diffusion timescale increases with rod length. The increase in residence time with swimming speed cannot be explained by such polydispersity, however, since longer rods tend to swim slightly slower than shorter rods.

Figure \ref{fig:time_on_teardrop} (inset) shows the distribution of mean residence times for rods swimming between $2-4 \, \mu$m/s. At larger residence times there is seen to be an approximately exponential decrease with a departure rate constant of $\sim 0.25$/s. This is consistent with our previous work showing an exponentially distributed residence time for swimmers orbiting $2.8 \mu$m radius spheres \cite{Takagi2014}, with a departure rate constant of $\sim 1.5$/s. The much lower departure rate observed here is consistent with the typically lower curvature of the tear-drop surface with which the rod interacts.  

\begin{figure}[t]
	\centering
	\includegraphics[width=0.49\textwidth]{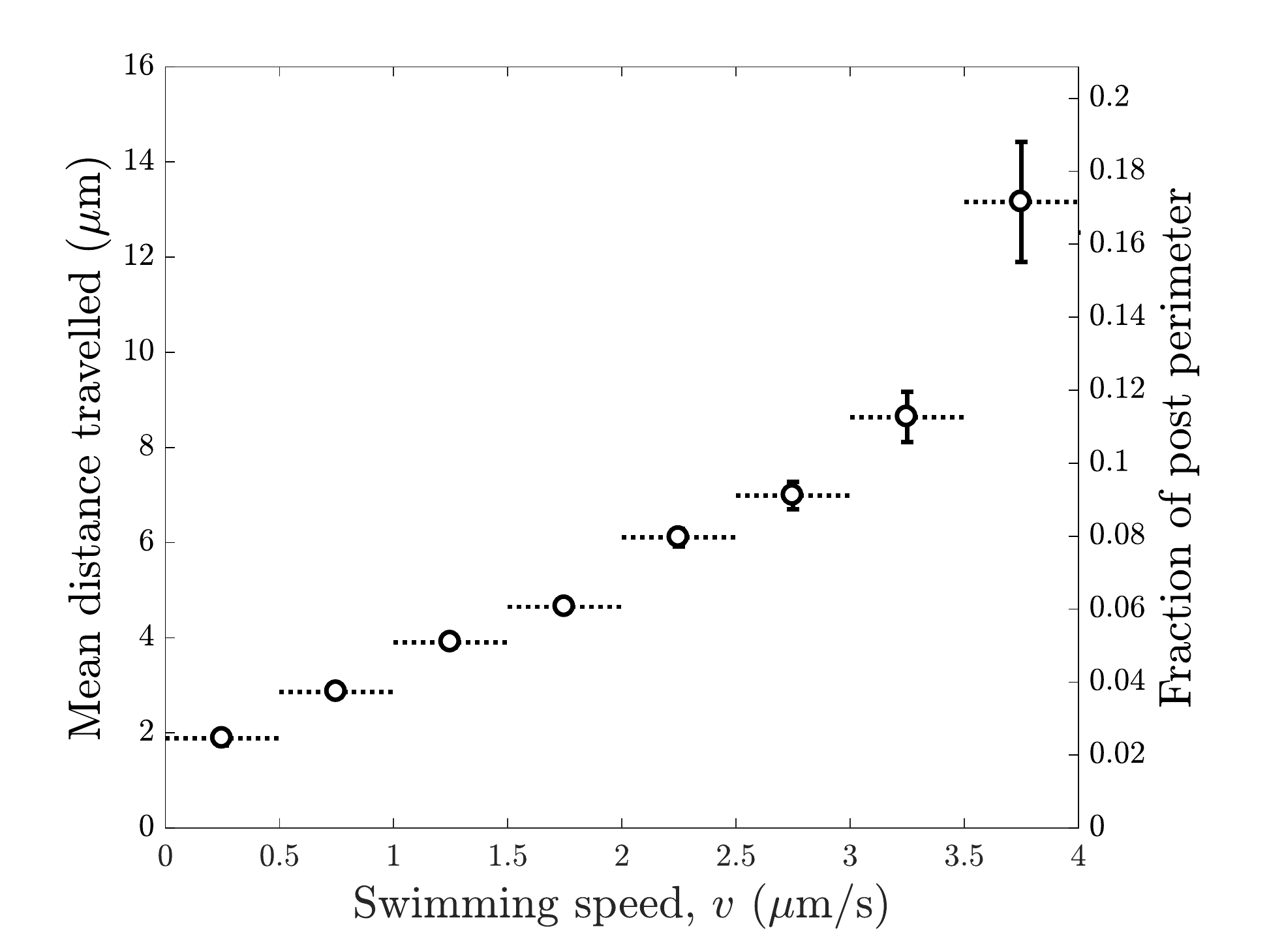}
	\caption{Mean distance travelled on teardrop-shaped posts as a function of mean swimming speed. All experiments had a H$_2$O$_2$ concentration of 10\%. Dotted horizontal lines correspond to range of rod velocities used to calculate the datapoint, vertical solid lines show the standard error of the mean.}
	\label{fig:distance_on_teardrop}
\end{figure}

Yet more important than the residence times of rods moving around the teardrop are the distances that they travel. The longer these distances the more likely a swimmer arrives at the teardrop tip where the likelihood of departure is highest. 
Figure \ref{fig:distance_on_teardrop} shows the mean distance travelled as a function of swimming speed. The mean distance travelled is calculated from the integrated path length, i.e.\ the sum of the distances moved between consecutive video frames, separated by $dt = 1/f$, where $f = 10$ fps is the frame rate.
 On average, microswimmers do not travel far around the teardrop perimeter. To give a sense of scale, a rod travelling at $4 \, \mu$m/s will, on average, only traverse $17\%$ of the total post perimeter. Nonetheless, at higher speeds there is the emergence of a superlinear increase in distance travelled. Attractive lubrication forces increase with swimmer speed \cite{Takagi2014}, which may underlie this superlinear increase in distance travelled. This feature is likely a contributor to the superlinear growth in both post and global bias seen in Figs. \ref{fig:single_teardrop_bias} and \ref{fig:array_bias}, respectively. 

Figures \ref{fig:time_on_teardrop} and \ref{fig:distance_on_teardrop} show that both residence time and distance traveled along the post increase with swimming speed, and it is tempting to interpret the ratio of these quantities as the average speed of the rods along the post. For example, the ratio of travel distance to residence time is $\sim 3\mu$m/s for rods of speed $\sim 1\mu$m/s, and this apparent discrepancy reflects the fact that the total distance traveled is influenced by both the deterministic velocity and by diffusion. Distance travelled, as measured by calculating the integrated path length between corresponding joining and leaving events, will be approximately $\sqrt{v^2 + 2Df} \times T$, where $f = 10$ fps is the frame rate of the video, $D$ is the translational diffusion coefficient, $v$ is the deterministic swimming speed, and $T$ is the residence time (Fig.\ \ref{fig:time_on_teardrop}). 

Other chemically-driven micro-swimmers show different behaviours in their interactions with boundaries. All experiments for Figs. \ref{fig:time_on_teardrop} and \ref{fig:distance_on_teardrop} were performed with a hydrogen peroxide concentration of 10\%. The increase in residence time with swimming speed at a constant hydrogen peroxide concentration contrasts with reports of Janus particle swimmers, which were found to have a residence time that varied with hydrogen peroxide concentration, but not with swimming speed \cite{Brown2016}. It is possible that this difference in the behaviour of the residence time is due to differences in the mechanisms that drive the motion. Janus particles swim by a mechanism that is currently not well-understood and have a swimming speed that saturates at relatively low hydrogen peroxide concentrations \cite{Brown2016} (around $3\%$), whereas the speed of bimetallic artificial swimmers increases linearly with hydrogen peroxide concentration for concentrations up to $25\%$ \cite{Takagi2013}. 

The approach described here for guiding and biasing microswimmer direction promises a protocol for separation of nanorods swimming at different speeds, using the dependence of bias on speed. The approach also suggests an effective and flexible strategy to move cargo-carrying swimmers through a microfluidic chip as well as a method for concentrating solutions of swimmers.

\section{Acknowledgements}

We thank J.\ Palacci and C.\ Chang for helpful discussions. This work was supported primarily by the Materials Research Science and Engineering Center (MRSEC) program of the National Science Foundation under award DMR-1420073 (TA, MSDW, MJS, MDW, XZ) and contributing support from the National Science Foundation under awards DMS-1463962 (MSDW, LR, MJS, JT, JZ) and DMS-1620331 (MJS). T.A.\ thanks the JSPS Postdoctoral Fellowships for Research Abroad for the financial support.

\bibliographystyle{rsc}
\bibliography{active_matter}

\end{document}